\documentclass[utf8]{frontiersSCNS} 

\usepackage{url,lineno,microtype,subcaption}
\usepackage[onehalfspacing]{setspace}
\usepackage{amsmath}
\usepackage{multirow}
\usepackage{float}

\def\keyFont{\fontsize{8}{11}\helveticabold }
\def\firstAuthorLast{Bakrania {et~al.}} 
\def\Authors{Mayur R. Bakrania\,$^{1,*}$, I. Jonathan Rae\,$^{1,2}$, Andrew P. Walsh\,$^{3}$, Daniel Verscharen\,$^{1,4}$ and Andy W. Smith\,$^{1}$}

\def\Address{$^{1}$Department of Space and Climate Physics, Mullard Space Science Laboratory, University College London, Dorking RH5 6NT, UK \\
$^{2}$Department of Mathematics, Physics and Electrical Engineering, University of Northumbria, Newcastle, UK \\
$^{3}$European Space Astronomy Centre, ESA, Urb. Villafranca del Castillo, Villanueva de la Cañada, Madrid, Spain \\
$^{4}$Space Science Center, University of New Hampshire, Durham, NH, USA}
\def\corrAuthor{Mayur R. Bakrania}

\def\corrEmail{mayur.bakrania@ucl.ac.uk}

\begin{document}
\onecolumn
\firstpage{1}

\title[Machine learning: classifying plasma regimes]{Using dimensionality reduction and clustering techniques to classify space plasma regimes} 

\author[\firstAuthorLast ]{\Authors} 
\address{} 
\correspondance{} 

\extraAuth{}

\maketitle

\begin{abstract}

Collisionless space plasma environments are typically characterised by distinct particle populations. Although moments of their velocity distribution functions help in distinguishing different plasma regimes, the distribution functions themselves provide more comprehensive information about the plasma state, especially at times when the distribution function includes non-thermal effects. Unlike moments, however, distribution functions are not easily characterised by a small number of parameters, making their classification more difficult to achieve. In order to perform this classification, we propose to distinguish between the different plasma regions by applying dimensionality reduction and clustering methods to electron distributions in pitch angle and energy space. We utilise four separate algorithms to achieve our plasma classifications: autoencoders, principal component analysis, mean shift, and agglomerative clustering. 
 
We test our classification algorithms by applying our scheme to data from the Cluster-PEACE instrument measured in the Earth’s magnetotail. Traditionally, it is thought that the Earth’s magnetotail is split into three different regions (the plasma sheet, the plasma sheet boundary layer, and the lobes), that are primarily defined by their plasma characteristics. Starting with the ECLAT database with associated classifications based on the plasma parameters, we identify 8 distinct groups of distributions, that are dependent upon significantly more complex plasma and field dynamics. By comparing the average distributions as well as the plasma and magnetic field parameters for each region, we relate several of the groups to different plasma sheet populations, and the rest we attribute to the plasma sheet boundary layer and the lobes. We find clear distinctions between each of our classified regions and the ECLAT results.
 
The automated classification of different regions in space plasma environments provides a useful tool to identify the physical processes governing particle populations in near-Earth space.  These tools are model independent, providing reproducible results without requiring the placement of arbitrary thresholds, limits or expert judgement. Similar methods could be used onboard spacecraft to reduce the dimensionality of distributions in order to optimise data collection and downlink resources in future missions.

\tiny
 \keyFont{ \section{Keywords:} space plasma environments, particle populations, distribution functions, dimensionality reduction, clustering techniques} 
\end{abstract}

\section{Introduction}

Particle populations in collisionless space plasma environments, such as the Earth’s magnetotail, are traditionally characterised by the moments of their distribution functions. 2D distribution functions in pitch angle and energy, however, provide the full picture of the state of each plasma environment, especially when non-thermal particle populations are present that are less easily characterised by a Maxwellian fit. These non-thermal plasma populations are ubiquitous across the solar system. They make crucial contributions to the bulk properties of a plasma, such as the temperature and collisionality \citep{cross-scale}. Magnetic reconnection, for example, heats non-thermal seed populations in both the diffusion and outflow regions, making them an important component of the overall energisation process \citep{energisation}. High-quality measurements and analysis of collisionless plasmas are consequently of key importance when attempting to understand these non-thermal populations.

Distribution functions, unlike moments, are not easily classified by a small number of parameters. We therefore propose to apply dimensionality reduction and clustering methods to particle distributions in pitch angle and energy space as a new method to distinguish between the different plasma regions. 2D distributions functions in pitch angle and energy are derived from full 3D distributions in velocity space based on the magnetic field direction and the assumption of gyrotropy of electrons. With these novel methods, we robustly classify variations in particle populations to a high temporal and spatial resolution, allowing us to better identify the physical processes governing particle populations in near-Earth space. Our method also has the advantage of being independent of the model applied, as these methods do not require prior assumptions of the distributions of each population. 

\subsection{Machine Learning Models}

In this section, we give a detailed account of the internal operations of each of the unsupervised machine learning algorithms used in our method. In unsupervised learning, algorithms discover the internal representations of the input data without requiring training on example output data. Dimensionality reduction is a specific type of unsupervised learning in which data in high-dimensional space is transformed to a meaningful representation in lower dimensional space. This transformation allows complex datasets, such as 2D pitch angle and energy distributions, to be characterised by analysis techniques (e.g. clustering algorithms) with much more computational efficiency. Our machine learning method utilises four separate algorithms: autoencoders \citep{autoencoders}, principal component analysis \citep[PCA,][]{pca}, mean shift \citep{mean-shift_origin}, and agglomerative clustering \citep{ac_clustering}. We obtain the autoencoder algorithm from the Keras library \citep{keras}, and the PCA, mean shift, and agglomerative clustering algorithms from the scikit-learn library \citep{scikit-learn}. 

We use the autoencoder to compress the data by a factor of 10 from a high-dimensional representation. We subsequently apply the PCA algorithm to further compress the data to a three-dimensional representation. The PCA algorithm has the advantage of being a lot cheaper computationally than an autoencoder, however the algorithm only captures variations that emerge from linear relationships in the data, while autoencoders also account for non-linear relationships in the dimensionality reduction process \citep{pca_advantage}. For this reason, we only utilise the PCA algorithm after the data have been compressed via an autoencoder. After compressing the data, we use the mean shift algorithm to inform us of how many populations are present in the data using this three-dimensional representation. While the mean shift algorithm provides us with this estimate of the requisite number of clusters, the algorithm is ineffective in constraining the shapes of the clusters to determine which population each data-point belongs to. Therefore, we use an agglomerative clustering algorithm to assign each data-point to one of the populations. 

\subsubsection{Autoencoders}
\label{sec:autoencoders}

Autoencoders are a particular class of unsupervised neural networks. They are trained to learn compressed representations of data by using a bottleneck layer which maps the input data to a lower dimensional space, and then subsequently reconstructing the original input. By minimising the `reconstruction error', or `loss', the autoencoder is able to retain the most important information in a representative compression and reconstruction of the data. As a result, autoencoders have applications in dimensionality reduction \citep[e.g.][]{autoencoders}, anomaly detection \citep[e.g.][]{outlier_AE} and noise filtering \citep[e.g.][]{de-noise}.

\begin{figure}[h!]
\begin{center}
\includegraphics[width=6cm]{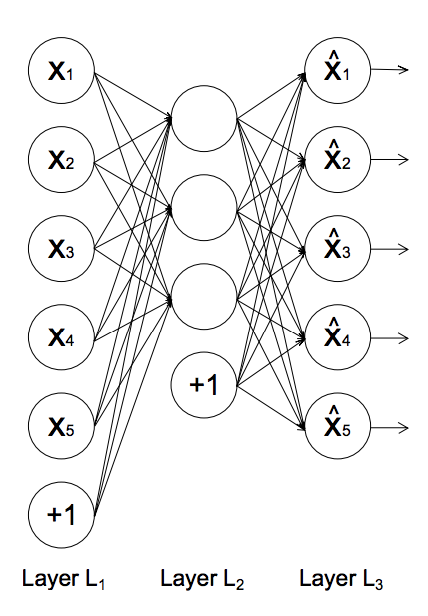}
\end{center}
\caption{The architecture of an autoencoder, adapted from \cite{diagram_AE}. Each circle represents a neuron corresponding to a data-point. Layer L\textsubscript{1} represents the input data, layer L\textsubscript{2} the encoded data in latent space, and layer L\textsubscript{3} the reconstructed data. The circles labelled `+1' are known as `bias units', which are parameters that are adjusted during training to improve the performance of the neural network.}
\label{fig:autoencoder}
\end{figure}

During training, an autoencoder runs two functions simultaneously. The first, called an `encoder', maps the input data, $\pmb{x}$, to the coded representation in latent space, $\pmb{z}$. The second function, called a `decoder', maps the compressed data, $\pmb{z}$, to a reconstruction of the input data, $\pmb{\hat{x}}$. The encoder, $E(\pmb{x})$, and decoder, $D(\pmb{z})$, are defined by the following deterministic posteriors:
\begin{equation}
\begin{split}
E(\pmb{x})=p(\pmb{z}|\pmb{x};\theta _E),\\
D(\pmb{z})=p(\pmb{\hat{x}}|\pmb{z};\theta _D),
\label{eq:autoencoder}
\end{split}
\end{equation}
where $\theta_{E}$ and $\theta_{D}$ are the trainable parameters of the encoder and decoder respectively. Figure \ref{fig:autoencoder} illustrates the standard architecture of an autoencoder.

In feed-forward neural networks, such as autoencoders, each neuron computes the following sum:
\begin{equation}
y = \sum_{i}w_{i}x_{i}+b,
\label{eq:activate_y}
\end{equation}
where $x_{i}$ represents the input from the previous layer, $w_{i}$ denotes the weights associated with the connections between neurons in different layers, and $b$ denotes the bias term associated with each layer (represented by the circles labelled `+1' in figure \ref{fig:autoencoder}). The number of neurons in each layer defines the dimension of the data representation in that layer. The output of each neuron, $f(y)$, is called the activation function. ReLU \citep[Rectified Linear Unit,][]{relu} is the most commonly used activation function due to its low computational cost \citep{cost_relu}. The function is described as:
\begin{equation}
f(y) = \max(0,y).
\label{eq:relu}
\end{equation}
The sigmoid activation function \citep{sigmoid} is also commonly used. It is defined by:
\begin{equation}
f(y) = \frac{1}{1+e ^{-y}}, 
\label{eq:sigmoid}
\end{equation}
where $y$ is defined in Equation \eqref{eq:activate_y}. Analysis of the use of various activation functions in the remit of plasma physics are given by \cite{outlier_AE}.

In order to improve the representation of the compressed data in layer L\textsubscript{2} and minimise the discrepancy between the input and reconstruction layer, the autoencoder adjusts the weights and biases by minimising a loss function through an optimiser (described below). The binary cross-entropy loss function \citep{cross-entropy} is typically used when the input data, $\pmb{x}$, are normalised to values between 0 and 1. The loss value, $c$, increases as the reconstruction data, $\pmb{\hat{x}}$, diverge from the input data. The loss function is defined as:
\begin{equation}
c = -[\pmb{x} \ln(\pmb{\hat{x}})+(1-\pmb{x})\ln(1-\pmb{\hat{x}})].
\label{eq:binary_ce}
\end{equation}
An overview of various loss functions is provided by \cite{loss_options}. Optimisers are used to ensure the autoencoder converges quickly to a minimum loss value by finding the optimum value of the weight, $w_{i}$, of each neuron. This is achieved by running multiple iterations with different weight values, known as gradient descent \citep{gradient_descent}. The weights are adjusted in each iteration, $t$, according to:
\begin{equation}
w_{t} = w_{t-1} - \alpha \frac{\partial c}{\partial \omega},
\label{eq:weight_descent}
\end{equation}
where $\partial c/\partial \omega$ is the gradient, which is a partial derivative of the loss value with respect to the weight. The learning rate, $\alpha$, updates all the weights simultaneously with respect to the gradient descent. This learning rate is randomly initialised between 0 and 1 by the algorithm. A low learning rate results in a slower convergence to the global minimum loss value. However a too high value for the learning rate impedes the gradient descent (Equation \ref{eq:weight_descent}) from converging on the optimum weights. The Adadelta optimiser \citep{Adadelta} is commonly used due to its rapid convergence to the minimum loss value and its ability to adapt the learning rate depending on each parameter. The optimiser updates each parameter, $\theta$, according to:
\begin{equation}
\triangle \theta_{t} = -\frac{\textrm{RMS}[\triangle \theta]_{t-1}}{\textrm{RMS}[g]_{t}}g_{t}, 
\label{eq:adadelta}
\end{equation}
where $\triangle\theta_{t}$ is the parameter update at the $t$-th iteration, $g_{t}$ is the gradient of the parameters at the $t$-th iteration, and RMS is the root mean square. An overview of the various optimisers is provided by \cite{optimiser_options}.

\vspace{1pt}
\subsubsection{Principal Component Analysis}
\label{sec:pca}

Principal component analysis is a statistical procedure that, as well as autoencoders, also reduces the dimensionality of input data. The algorithm achieves this by transforming the input data from a large number of correlated variables to a smaller number of uncorrelated variables, known as principal components. These principal components account for most of the variation in the original input data, making them a useful tool in feature extraction.

Before the procedure, the original data, $\pmb{X_{0}}$, are represented by a $(n \times Q)$ matrix, where $n$ is the number of observations and $Q$ is the number of variables (also called dimensions). In the first step, the algorithm scales and centres the data:
\begin{equation}
\pmb{X} = (\pmb{X_{0}}-\pmb{\bar{X_{0}}})\pmb{D}^{-1},
\label{eq:pca_scale}
\end{equation}
where $\pmb{\bar{X_{0}}}$ contains the means of each of the variables, and $\pmb{D}$ is a diagonal matrix that contains the scaling coefficient of each variable. Typically, $D_{ii} = \sigma_{i}$ where $\sigma_{i}$ is the standard deviation of variable with index $i$ \citep{pca_matrix}. The algorithm then uses $\pmb{X}$ to calculate the covariance matrix:
\begin{equation}
\pmb{C_{X}} = \frac{1}{n-1}\pmb{X}^{T}\pmb{X},
\label{eq:covariance}
\end{equation}
which measures the correlation between the different variables. The principal components are calculated as the eigenvectors, $\pmb{A}$, of the covariance matrix:
\begin{equation}
\pmb{C_{X}} = \pmb{A}\pmb{L}\pmb{A}^{T},
\label{eq:cov_eigen}
\end{equation}
where $\pmb{L}$ is a diagonal matrix containing the eigenvalues associated with $\pmb{A}$. These principal components are ordered in decreasing order, whereby the first principal components account for most of the variation in the input data. These input data are finally projected into the principal component space according to:
\begin{equation}
\pmb{Z} = \pmb{X}\pmb{A},
\label{eq:output_pca}
\end{equation}
where $\pmb{Z}$ represents the output data containing the principal component scores. The dimensionality of these output data are determined by the number of principal components used.

\subsubsection{Mean Shift}
\label{sec:mean_shift}

The mean shift algorithm is a non-parametric clustering technique that is used for locating the maxima of a density function in a sample space. The algorithm aims to discover the number of clusters within a dataset, meaning no prior knowledge of the number of clusters is necessary. 

For a dataset containing $n$ data-points $\pmb{x}_{i}$, the algorithm starts finding each maximum of the dataset's density function by randomly choosing a data-point to be the mean of the distribution, $\pmb{x}$. The algorithm then uses a kernel function, $K$, to determine the weights of the nearby data-points for re-estimating the mean. The variable $h$ is the width of the kernel window. Typically, a Gaussian kernel, $k$, is used:
\begin{equation}
K \left (\frac{\pmb{x}-\pmb{x}_{i}}{h} \right ) = c_{k}k \left (\left \| \frac{\pmb{x}-\pmb{x}_{i}}{h} \right \|^{2} \right) = \exp \left(-c_{k}\left \| \frac{\pmb{x}-\pmb{x}_{i}}{h} \right \|^{2} \right),
\label{eq:gauss_kernal}
\end{equation}
where $c_{k}$ is the normalising constant. With the kernel function, the multivariate kernel density estimator is obtained:
\begin{equation}
f(\pmb{x})=\frac{1}{nh^{d}} \sum_{i=1}^{n}K \left (\frac{\pmb{x}-\pmb{x}_{i}}{h} \right),
\label{eq:density_estimator}
\end{equation}
where $d$ is the dimensionality of the dataset. The gradient of the density estimator is then:
\begin{equation}
\begin{split}
\pmb{\triangledown}f(\pmb{x}) &= \frac{2c_{k}}{nh^{d+2}}\sum_{i=1}^{n}(\pmb{x}_{i}-\pmb{x})g\left ( \left \| \frac{\pmb{x}-\pmb{x}_{i}}{h} \right \|^{2} \right ) \\
& = \frac{2c_{k}}{nh^{d+2}}\left [ \sum_{i=1}^{n} g\left ( \left \| \frac{\pmb{x}-\pmb{x}_{i}}{h} \right \|^{2} \right ) \right ]\pmb{m}_{h}(\pmb{x}),
\label{eq:gradient_density_estimator}
\end{split}
\end{equation}
where $g(\pmb{x}) = -k'(\pmb{x})$. The first term is proportional to the density estimate at $\pmb{x}$, and the second term, $\pmb{m}_{h}(\pmb{x})$, is:
\begin{equation}
\pmb{m}_{h}(\pmb{x}) = \frac{\sum_{i=1}^{n}\pmb{x}_{i}g\left ( \left \| \frac{\pmb{x}-\pmb{x}_{i}}{h} \right \|^{2}  \right )}{\sum_{i=1}^{n} g\left ( \left \| \frac{\pmb{x}-\pmb{x}_{i}}{h} \right \|^{2}  \right )} -\pmb{x},
\label{eq:ms_vector}
\end{equation}
which is the mean shift vector and points towards the direction of the maximum increase in density. The mean shift algorithm therefore iterates between calculating the mean shift vector, $\pmb{m}_{h}(\pmb{x}^{t})$, and translating the kernel window:
\begin{equation}
\pmb{x}^{t+1} = \pmb{x}^{t} + \pmb{m}_{h}(\pmb{x}^{t}), 
\label{eq:ms_iteration}
\end{equation}
where $t$ is the iteration step. Once the window has converged to a point in feature space where the density function gradient is zero, the algorithm carries out the same procedure with a new window until all data-points have been assigned to a maximum in the density function.

\subsubsection{Agglomerative Clustering}
\label{sec:agglomerative_clustering}

Agglomerative clustering is a type of hierarchical clustering that uses a `bottom-up' approach, whereby each data-point is first assigned a different cluster. Then pairs of similar clusters are merged until the specified number of clusters has been reached. During each recursive step, the agglomerative clustering algorithm combines clusters typically using Ward's criterion \citep{Ward_criterion}, which finds pairs of clusters that lead to the smallest increase in the total intra-cluster variance after merging. The increase is measured by a squared Euclidean distance metric:
\begin{equation}
d_{ij} = d(C_{i},C_{j}) = \left \| C_{i} - C_{j} \right \|^{2}, 
\label{eq:euclidean}
\end{equation}
where $C_{i}$ represents a cluster with index $i$. The algorithm implements Ward's criterion using the Lance-Williams formula \citep{lance-williams}:
\begin{equation}
\begin{split}
d(C_{i} \cup C_{j},C_{k}) &= \frac{n_{i}+n_{k}}{n_{i}+n_{j}+n_{k}}d(C_{i},C_{k}) \\
&+ \frac{n_{j}+n_{k}}{n_{i}+n_{j}+n_{k}}d(C_{j},C_{k}) - \frac{n_{k}}{n_{i}+n_{j}+n_{k}}d(C_{i},C_{j}), 
\label{eq:lance-williams}
\end{split}
\end{equation}
where $C_{i}$, $C_{j}$, and $C_{k}$ are disjoint clusters with sizes $n_{i}$, $n_{j}$, and $n_{k}$, and $d(C_{i} \cup C_{j},C_{k})$ is the squared Euclidean distance between the new cluster $C_{i} \cup C_{j}$ and $C_{k}$. The clustering algorithm uses Equation \eqref{eq:lance-williams} to find the optimal pair of clusters to merge.

\subsection{The Magnetotail}

We use electron data from the magnetotail in order to test the effectiveness of our method. The magnetotail is traditionally divided into three different regions: the plasma sheet (PS), the plasma sheet boundary layer (PSBL), and the lobes \citep{magnetotail_regions}. These regions are defined by their plasma and magnetic field characteristics. The low temperature ($\sim$85 eV) outermost northern and southern lobes are on open magnetic field lines which results in a much lower plasma density of $\sim$0.01 cm$^{-3}$ \citep{Lui_87}. The plasma sheet boundary layer exists on the reconnected magnetic field lines. This region forms the transition region in between the plasma sheet and the lobes, and is characterised by a population of field-aligned particles and a plasma $\beta$, which is the ratio of the plasma pressure to the magnetic pressure, of $\sim$0.1 \citep{Lui_87}. 

The innermost plasma sheet typically contains a comparatively hot ($\sim$4250 eV) and isotropic plasma with a relatively high particle density of $\sim$0.01 cm$^{-3}$. At the centre of the plasma sheet is the thin neutral current sheet, which is characterised by a relatively high plasma $\beta$ of $\sim$10, and a magnetic field strength of near zero \citep{Lui_87}. Although isotropic electron pitch angle distributions (PADs) are the most dominant in the plasma sheet, many cases of pitch angle anisotropy have also been found \citep[e.g.][]{walsh_2013,Artemyev_2014,Liu_pads}. These intervals correspond to a colder and denser electron population and are linked to: cold anisotropic ionospheric outflows \citep{walsh_2013}, and a penetration of cold electrons from the magnetosheath near the flanks \citep{Artemyev_2014}.

\section{Method and Application}

In this section, we detail the steps required to classify different regions within a space plasma environment using machine learning techniques. As an example, we classify Cluster-PEACE \citep[Plasma Electron And Current Experiment,][]{peace,faz_peace} data \citep{csa} from the Earth's magnetotail to showcase our method, as this allows us to compare to the Cluster-ECLAT \citep{eclat} database for evaluation. The same method, however, can be applied to any plasma regime where energy and pitch angle measurements are available. Our steps are as follows:

\begin{enumerate}
    \item \textbf{Data preparation:} We obtain the Cluster-PEACE data from different magnetotail regions based on the Cluster-ECLAT database, and prepare the data for testing.
    \item \textbf{Reducing dimensionality:} We build our autoencoder and use the encoder part to reduce the dimensionality of each pitch angle and energy distribution by a factor of 10. We use a PCA algorithm to further compress each distribution to a set of coordinates in 3D space.
    \item \textbf{Clustering:} We apply the mean-shift algorithm to determine how many clusters exist within the compressed magnetotail electron data, and use an agglomerative clustering algorithm to separate the compressed dataset into this number of clusters. This allows us to determine how many plasma regimes exist within the overall dataset.
    \item \textbf{Evaluation:} We estimate the probabilities of the agglomerative clustering labels and compare our clustering results to the original ECLAT labels in order to evaluate our method.
\end{enumerate}

\subsection{Data Preparation}

\begin{figure}[h!]
\begin{center}
\includegraphics[width=11cm]{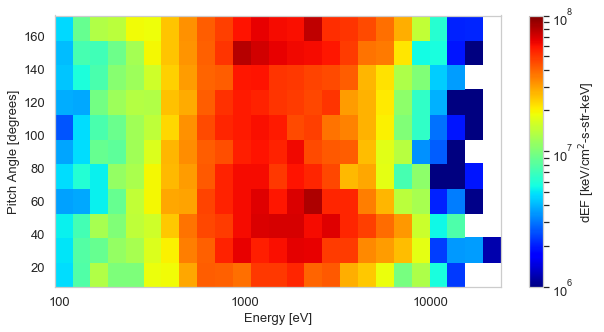}
\end{center}
\caption{An example two-dimensional electron differential energy flux distribution, as a function of pitch angle (degrees) and energy (eV), measured by the Cluster-PEACE instrument in the magnetotail across a 4 second window (09:51:23-09:51:27 on 13/10/2003).}
\label{fig:t7164_2d_tail}
\end{figure}

We prepare PEACE instrument data from the Cluster mission's C4 spacecraft \citep{escoubet_2001} to test and present our method. The Cluster mission comprises of four spacecraft, each spinning at a rate of 4 s\textsuperscript{-1}. The PEACE data have a 4 s time resolution and are constructed from two instantaneous pitch angle distribution measurements per spin. Each of our distributions is a two-dimensional differential energy flux product containing twelve 15$^{\circ}$ wide pitch angle bins and 26 energy bins, spaced logarithmically between 93 eV and 24 keV. The dimensionality of each distribution is 312 (12$\times$26). We normalise the differential energy flux linearly between 0 and 1 based on the maximum flux value in the dataset. An example of an individual differential energy flux distribution used in our analysis is shown in figure \ref{fig:t7164_2d_tail}. We correct for spacecraft potential with measurements from the Cluster-EFW instrument \citep{gustafsson_efw} and corrections (19\% increase) according to the results of \cite{cully}.

The ECLAT dataset consists of a detailed list of plasma regions encountered by each of the four Cluster spacecraft in the nightside magnetosphere. The dataset is available from July to October during the years 2001-2009. Using plasma and magnetic field moments from the PEACE, CIS \citep{reme_2001} and FGM \citep{cluster_fgm} instruments, the dataset provides a list of (inner and outer) plasma sheet, boundary layer, and lobe times. These regions are identified based on the plasma $\beta$, the magnetic field measurements, and the current density vectors. A comprehensive account of the ECLAT identification routine for each plasma region is provided by \cite{eclat}. To ensure that we test our method on a large number of data from each of the magnetotail regions ($>$50 000 samples), we obtain PEACE data from times when the C4 spacecraft has spent at least 1 hour in each region, according to ECLAT. 

\subsection{Reducing Dimensionality}

After preparing the dataset to include a series of $>$50 000 time intervals, each with its associated 2D pitch angle and energy distributions (e.g. figure \ref{fig:t7164_2d_tail}), the first step towards reducing the dataset's dimensionality is to build a suitable autoencoder (described in Section \ref{sec:autoencoders}). We construct our autoencoder using the Keras library. This step requires defining the number of neurons in each layer. The input and reconstruction layer should have the same number, which is equal to the dimensionality of the original dataset (312 for each time interval in this example). The middle encoded layer typically contains a compressed representation of the data that is by a factor of 10 smaller than the input data \citep{autoencoders}. We therefore specify our encoded layer to contain 32 neurons. The next step involves specifying the activation function for the neurons in the first and middle layers. We use the standard ReLU activation function \citep{relu} in the encoder part of our autoencoder and the sigmoid activation function \citep{sigmoid} in the decoder part, as this function is used to normalise the output between 0 and 1. 

The next step defines which loss function and optimiser the autoencoder uses in order to representatively compress and reconstruct the input data. As we use normalised output data, we choose the standard binary cross-entropy loss function \citep{cross-entropy}. In terms of the optimiser, we utilise the Adadelta optimiser \citep{Adadelta} due to its speed and versatility. All of the activation functions, loss functions, and optimisers are available in the Keras library.

In the next step, we set the hyperparameters used for training the autoencoder. These hyperparameters include: the number of epochs, the batch size, and the validation split ratio. The number of epochs represents the number of training iterations undergone by the autoencoder, with the weights and biases updated at each iteration. The batch size defines the number of samples that are propagated through the network at each iteration. It is equal to $2^{n}$, where $n$ is a positive integer. The batch size (256 in our case) is ideally set as close to the dimensionality of the input data as possible. The validation split ratio determines the percentage of the input data that should remain `unseen' by the autoencoder in order to verify that the algorithm is not overfitting the remaining training data. We set the validation split ratio to 1/12, which is commonly used for large datasets \citep{validation_split}. At each iteration, a training loss value and a validation loss value are produced, which are determined by the binary cross-entropy loss function. Both of these values converge to their minima after a certain number of iterations, at which point the autoencoder cannot be optimised to the input data any further. Loss values $<$0.01 are typically considered ideal \citep{accurate_loss}. 

After retrieving the compressed representation of the input data from the encoding layer (with a dimensionality of 32 in our case), we apply a PCA algorithm (see Section \ref{sec:pca}) to the compressed data to reduce the dimensionality to 3. We obtain the PCA algorithm from the scikit-learn library. We set the output dimensionality of the PCA algorithm to 3 as the following clustering algorithms used in this method are computationally expensive and their performance scales poorly with increasing dimensionality \citep{mean-shift,ac_clustering}. Setting the dimensionality to 3 has the added benefit that the clusters can be visualised.

\subsection{Clustering}

Once the dimensionality reduction stage has taken place and each pitch angle and energy distribution is represented by 3 PCA values, we use clustering algorithms to separate the dataset into the different particle populations. To first determine how many populations exist within the dataset (8 in our case), we apply a mean shift clustering algorithm (see Section \ref{sec:mean_shift}) to the data to find the number of maxima, $n_{c}$, in the distribution of data-points. We obtain the mean shift algorithm from the scikit-learn library. We set the bandwidth, represented by $h$ in Equation \eqref{eq:ms_vector}, to 1, which we find optimises the time taken for the algorithm to converge on the maxima in the density distribution. 

After determining the number of clusters in the dataset, we use an agglomerative clustering algorithm (see Section \ref{sec:agglomerative_clustering}) to assign each data-point to one of the $n_{c}$ clusters. We obtain the agglomerative clustering algorithm from the scikit-learn library and instantiate the algorithm by specifying the number of clusters, $n_{c}$, before applying it to the compressed dataset. Assigning several clusters to a large dataset with 3 dimensions is a computationally expensive task, however we find the agglomerative clustering algorithm converges relatively quickly in comparison to other clustering algorithms. A further advantage of the hierarchical clustering procedure, used in the agglomerative clustering algorithm, is that data-points belonging to a single non-spherical structure in the 3-dimensional parameter space are not incorrectly separated into different clusters, unlike the more widely used K-means algorithm \citep{k-means}. 

\begin{figure}[h!]
\begin{center}
\includegraphics[width=17cm]{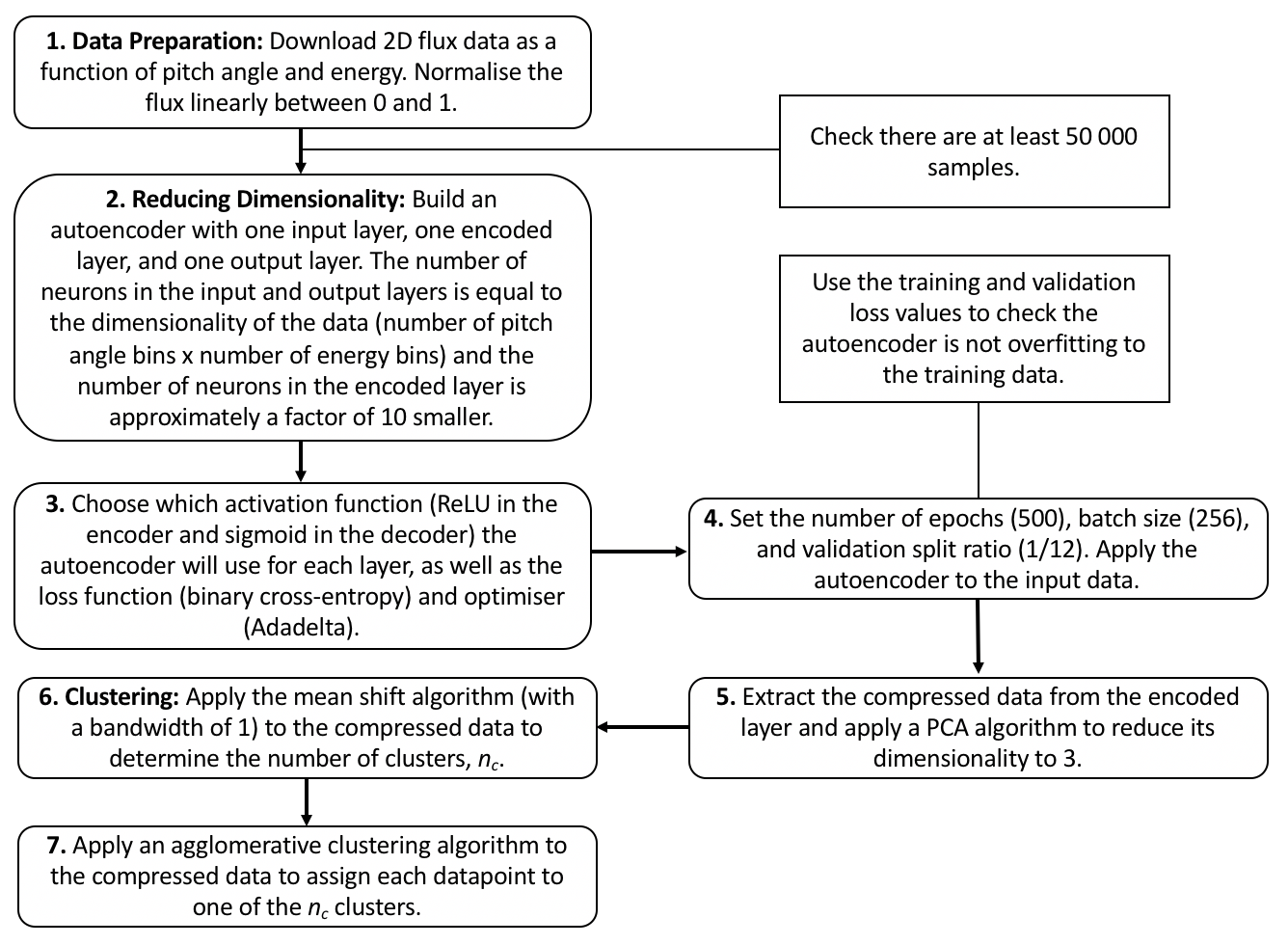}
\end{center}
\caption{Flow diagram illustrating the steps we take to reduce the dimensionality of the dataset and subsequently apply clustering algorithms to characterise the different populations. Our choices for the functions and input parameters necessary to train the autoencoder are shown in brackets in steps 3 and 4.}
\label{fig:flow_diagram}
\end{figure}

Figure \ref{fig:flow_diagram} contains a flow diagram detailing our method.

\subsection{Evaluation}
\label{sec:Evaluation}

\begin{figure}[h!]
\begin{center}
\includegraphics[width=9cm]{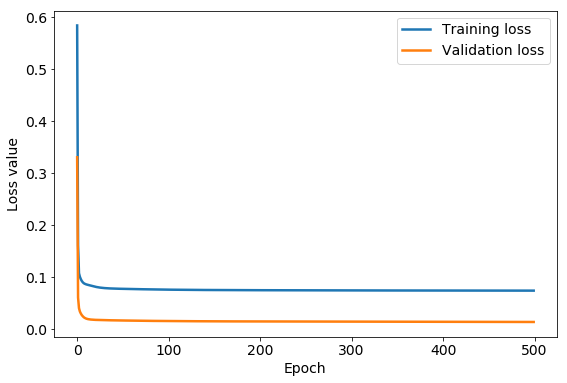}
\end{center}
\caption{The evolution of the training loss value and validation loss value as the autoencoder iterates through 500 steps (epochs).}
\label{fig:loss_curve}
\end{figure}

Figure \ref{fig:loss_curve} shows the training and validation loss values associated with each iteration during the training of our autoencoder. We use this graph to check if the autoencoder is overfitting to the training data, which is evident if the training loss starts to decrease more rapidly than the validation loss. In this case, our autoencoder is not overfitting at any iteration during training. Figure \ref{fig:loss_curve} shows that both the loss values start to rapidly level off in less than 100 epochs. Both loss values, however, continue to decrease, with the training loss value converging to 0.0743 after 444 iterations, and the validation loss value converging to 0.0140 after 485 iterations. We therefore set the number of epochs to 500. As both loss values are lower than 0.01, we conclude the autoencoder is accurately reconstructing both sets of input data, assuring us that the encoded data with a lower dimensionality is representative of the original 2D distribution functions. The lower validation loss than training loss in figure \ref{fig:loss_curve} indicates the presence of anomalous data in the training set that is not represented in the validation set. We discuss this anomalous data later in this section.

\begin{figure}[h!]
\begin{center}
\includegraphics[width=13cm]{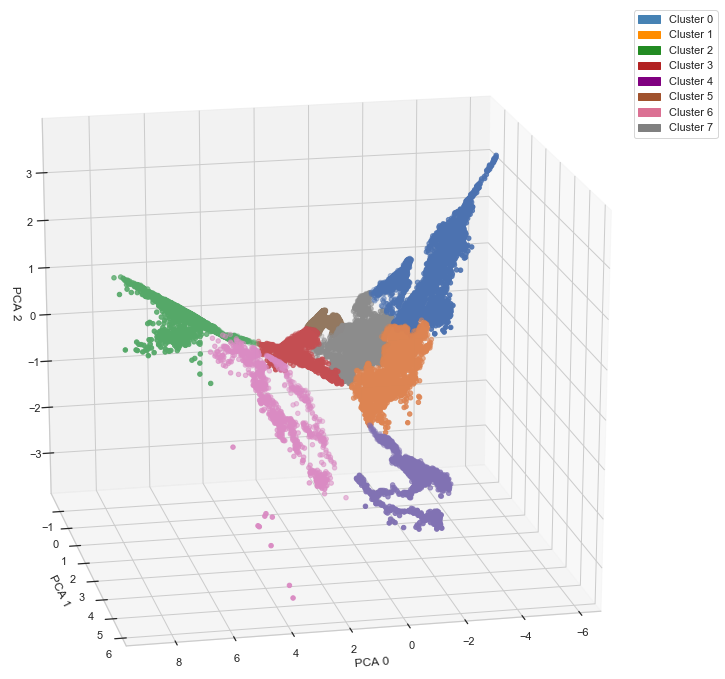}
\end{center}
\caption{Three-dimensional representation of the magnetotail data after undergoing dimensionality reduction via an autoencoder and PCA algorithm. The colours represent the clustering results from the agglomerative clustering algorithm.}
\label{fig:ac_3d_rotate_aligned}
\end{figure}

Figure \ref{fig:ac_3d_rotate_aligned} shows the result of applying the agglomerative clustering algorithm to the compressed magnetotail electron data after the implementation of the autoencoder and PCA algorithms. The 3-dimensional representation shows that the clustering algorithm is able to assign data-points of varying PCA values to the same cluster if they belong to the same complex non-spherical structure, e.g. clusters 0, 4, and 6. The clustering algorithm is able to form clear boundaries between clusters with adjacent PCA values, e.g. between clusters 0, 1, and 7, with no mixing of cluster labels on either side of the boundaries. The clustering algorithm locates the boundaries by finding areas with a low density of data-points in comparison to the centres of the clusters. 

\begin{figure}[h!]
\begin{center}
\includegraphics[width=17.5cm]{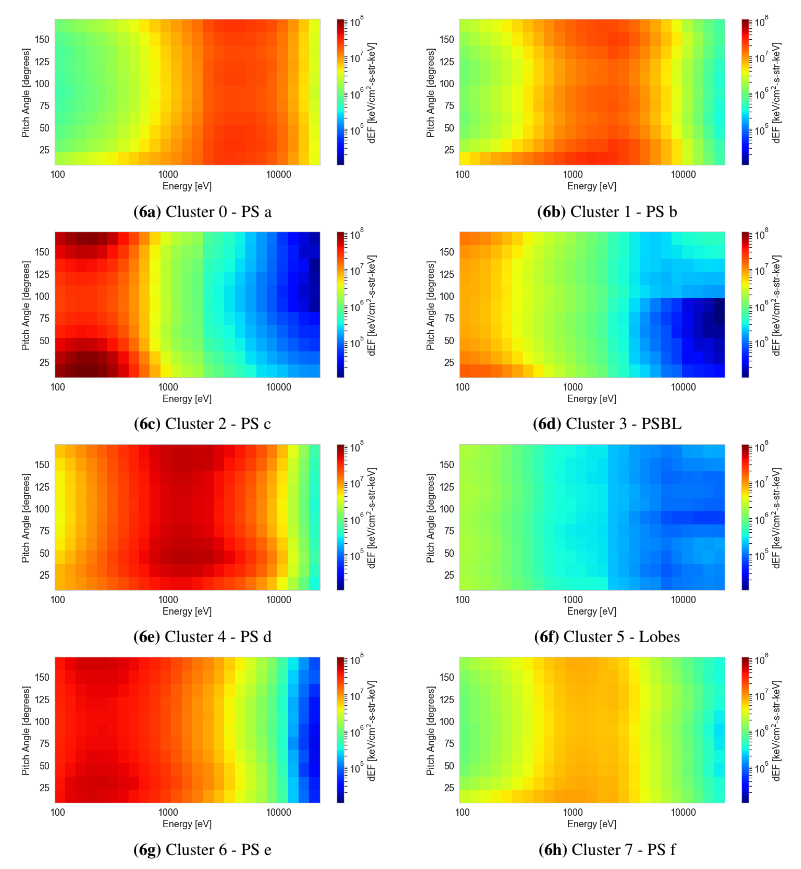}
\end{center}
\caption{Average electron differential energy flux distributions as a function of pitch angle and energy for each of the eight clusters classified by the agglomerative clustering algorithm. Each cluster is assigned a magnetotail region based on our interpretation of their plasma and magnetic field parameters.}
\label{fig:cluster_all_regions}
\end{figure}

Figure \ref{fig:cluster_all_regions} shows the results of averaging the 2D differential energy flux distributions in pitch angle and energy space for each of the 8 clusters. Using moments data collected by the PEACE, FGM, and CIS instruments, we compare the proton plasma $\beta$s, electron densities and temperatures, and magnetic field strengths to the average 2D distribution of each cluster. This process allows us to verify the consistency of the clustering method and provide general region classifications in order to make comparisons with the ECLAT labels. Our classifications (shown in the captions below each sub-figure) are produced with the aid of previous analyses of electron pitch angle distributions \citep[e.g.][]{walsh_2011,Artemyev_2014} and the plasma and magnetic field parameters \citep[e.g.][]{Lui_87,Artemyev_2014} in the magnetotail.

The individual sub-figures in figure \ref{fig:cluster_all_regions} display large differences in the average electron 2D pitch angle and energy distributions. Each average distribution differs by either: the energy of the peak flux, the peak value of the flux, or the amount of pitch angle anisotropy, i.e. the difference in flux between the parallel and perpendicular magnetic field direction. The lack of identical average distributions amongst the clusters shows the mean shift algorithm has not overestimated the number of clusters. By observing the individual 2D distributions within each cluster, we see a distinct lack of intra-cluster variance, showing the mean shift algorithm does not underestimate the number of clusters. 

A limitation of using the agglomerative clustering algorithm is that outliers or anomalous data are not differentiated from the main clusters. Clustering a sizeable number of outliers with the main populations can lead to ambiguity in the defining characteristics of each population, reducing the robustness of our method. In our case, figure \ref{fig:ac_3d_rotate_aligned} shows only 9 data points, within cluster 6, that are disconnected from the main structure of cluster 6 due to their distinct PCA values. We observe similar phenomenon to a lesser extent with cluster 2. To counteract this issue, we perform an outlier detection procedure using the reconstructed output of the autoencoder. By calculating the mean square error (MSE) between each input data-point and its reconstructed output, we isolate outliers in the dataset from the agglomerative clustering analysis based on their large MSE values, in comparison to 99.95\% of the data-points. During training, the autoencoder learns the latent space representation that defines the key characteristics of the bulk populations present in the dataset. The most relevant features of an anomalous particle distribution are not present in this subspace, resulting in a large mean square error between the reconstructed data, which lacks these important features, and the original data. This technique effectively identifies the 9 obvious outliers observable by eye in figure \ref{fig:ac_3d_rotate_aligned}, along with 6 from cluster 2 and 5 from cluster 1.

We use Gaussian mixture models \citep[GMMs,][]{GMMs} to establish the probabilities of each of the data-points belonging to the clusters they have been assigned to by the agglomerative clustering algorithm, providing useful information on the uncertainty associated with our region classification method. We obtain the GMM from the scikit-learn library. For each data-point, $x_{i}$, a GMM fits a normal distribution, $\mathcal{N}$, to each cluster and computes the sum of probabilities as:
\begin{equation}
p(x_{i}) = \sum^{k}_{j=1}\phi_{j}\mathcal{N}(x_{i};\mu_{j},\tau_{j}) = 1, 
\label{eq:GMMs}
\end{equation}
where $\mu_{j}$ and $\tau_{j}$ are the mean and covariance of the normal distribution belonging to cluster $j$, and $\phi_{j}$ is the mixing coefficient which represents the weight of Gaussian $j$ and is calculated by the Expectation-Maximisation (EM) algorithm \citep{EM_algorithm}. A complete description of GMMs and the EM algorithm is provided by \cite{Dupuis_2020}.

\begin{figure}[h!]
\begin{center}
\includegraphics[width=11cm]{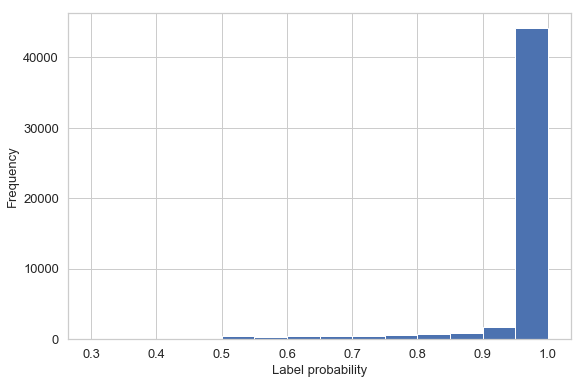}
\end{center}
\caption{Histogram showing the probabilities, generated by GMMs, that the data-points belong to the cluster assigned to them by the agglomerative clustering algorithm.}
\label{fig:gmm_graph}
\end{figure}

Figure \ref{fig:gmm_graph} shows a histogram of the probabilities, calculated by the GMM, associated with each data-point belonging to the cluster it is assigned to by the agglomerative clustering algorithm. More than 92\% of the data-points have a probability of over 0.9, and $<$1\% of the data-points have a probability of $<$0.5. This indicates a high certainty in our clustering method and validates the high precision in our region classifications. Further investigations of the data-points with associated probabilities of $<$0.5 show that these data-points exist on the boundary between clusters 0 and 1, i.e. two plasma sheet populations that differ by temperature. This illustrates a small limitation in the agglomerative clustering method when distinguishing between relatively similar plasma regimes.

Table \ref{tab:parameters} shows the median and upper and lower quartiles of the electron density, electron temperature, magnetic field, and ion plasma $\beta$ for each of the 8 clusters designated by our agglomerative clustering algorithm. None of the 8 clusters have comparable median and quartile values across all four of the chosen parameters. Certain pairs of clusters exhibit similarities in the median and quartile values for one or two of the four parameters, e.g. clusters 0 and 4 exhibit similar electron densities and magnetic field strengths, and clusters 3 and 6 exhibit similar magnetic field strengths. However there are large differences in the values of the remaining parameters for these pairs of clusters. These results show that clear differences in the 2D pitch angle and energy distributions (see figure \ref{fig:cluster_all_regions}) can translate into distinctions between certain but not all plasma parameter measurements, providing a strong indicator that full 2D distributions can effectively be used to distinguish between similar particle populations. Regarding the ECLAT classifications, which are based on magnetic field and plasma $\beta$ measurements, certain pairs of clusters exhibit a similar range of values in both of these measurements, e.g. clusters 0 and 4 and clusters 1 and 7. As the majority of data-points for all of these clusters are considered the same plasma sheet population by ECLAT (see table \ref{tab:contingency_table}), we conclude that using a limited number of parameters to provide classifications overlooks distinctions between different populations and incorrectly groups them into the same category.

\begin{table}[h!]
\caption{Comparisons of the median, Q2, and upper, Q3, and lower, Q1, quartile values of the electron density $n_{e}$, electron temperature $T_{e}$, magnetic field $\left | \boldsymbol{B} \right |$, and plasma $\beta$ associated with each of the 8 clusters. The AC labels 0, 1, 2, 4, 6, and 7 belong to the plasma sheet, according to ECLAT, 3 belongs to the plasma sheet boundary layer, and 5 belongs to the lobes.}
\centering
\resizebox{\columnwidth}{!}{%
\begin{tabular}{c|ccc|ccc|ccc|ccc}
\multirow{2}{*}{AC labels} & \multicolumn{3}{c}{$n_{e}$ (cm\textsuperscript{-3})} \vline & \multicolumn{3}{c}{$T_{e}$ (eV)} \vline & \multicolumn{3}{c}{$\left | \boldsymbol{B} \right |$ (nT)} \vline & \multicolumn{3}{c}{plasma $\beta$} \\ \cline{2-13}
                           & Q1         & Q2         & Q3    & Q1         & Q2         & Q3   & Q1         & Q2         & Q3  & Q1         & Q2         & Q3     \\ \hline 
0                          & 0.21      & 0.22         & 0.23     & 2057.26      & 2515.28         & 2913.33  & 32.45      & 35.09         & 37.86   & 3.54      & 4.33         & 5.92    \\
1                          & 0.12      & 0.14         & 0.19   & 1487.67      & 1838.97         & 2258.43    & 10.56      & 13.65         & 19.42   & 15.31      & 24.54       & 39.75      \\
2                        & 1.08      & 1.18      & 1.30    & 106.44      & 114.19        & 123.33  & 20.87      & 22.67         & 24.25  & 0.80      & 0.99         & 1.17   \\
3    &  0.21  &  0.25  &  0.28  &  79.21  &  83.63  &  93.00  &  16.64  &  35.07  &  39.76  &  0.28  &  0.39  &  0.60  \\
4                       &  0.22 &	0.28 &	0.82 &	783.33 &	879.77 &	997.85 &	24.58 &	39.10 &	41.93 &	1.23 &	1.52 &	6.29   \\
5              &   0.01 &	0.02 &	0.03 &	116.63 &	170.57 &	252.85 &	32.97 &	41.34 &	49.00 &	0.00 &	0.06 &	0.27       \\
6            &   1.29 &	1.49 &	1.65 &	164.41 &	214.64 &	269.16 &	41.65 &	44.74 &	46.56 &	0.30 &	0.43 &	0.60       \\
7                     &   0.08 &	0.10 &	0.13 &	669.64 &	882.30 &	1217.77 &	5.42 &	17.16 &	25.95 &	3.69 &	9.05 &	128.11
\end{tabular}
}
\label{tab:parameters}
\end{table}

Table \ref{tab:contingency_table} shows our comparison between the eight agglomerative clustering (AC) labels and the region names given in the ECLAT database, for the magnetotail data used in our example.

\begin{table}[h!]
\caption{Contingency table comparing the agglomerative clustering (AC) labels of the magnetotail electron data to the original ECLAT labels (0 = PS, 1 = PSBL, and 2 = lobes). The AC labels are the same as in table \ref{tab:parameters}.}
\centering
\begin{tabular}{c|ccc}
\multirow{2}{*}{AC labels} & \multicolumn{3}{c}{ECLAT labels} \\ \cline{2-4}
                           & 0         & 1         & 2         \\ \hline 
0                          & 6549      & 0         & 0         \\
1                          & 3074      & 0         & 0         \\
2                          & 5092      & 0         & 0         \\
3                          & 1590      & 4188      & 0         \\
4                          & 2097      & 0         & 0         \\
5                          & 156       & 2228      & 15641     \\
6                          & 1029      & 0         & 0         \\
7                          & 7020      & 1057      & 0        
\end{tabular}
\label{tab:contingency_table}
\end{table}

In table \ref{tab:contingency_table}, there is some disagreement with three of our clusters, namely AC labels 3, 5, and 7, which correspond to the plasma sheet boundary layer, the lobes, and a plasma sheet population respectively. However for each of these clusters, the majority of labels are in agreement with the ECLAT regions (72.4\%, 86.8\%, and 86.9\% for AC clusters 3, 5, and 7 respectively). For AC labels 0, 1, 2, 4, and 6, which represent various other populations within the plasma sheet, there is 100\% agreement with the ECLAT label 0, which denotes the plasma sheet. By using this method to characterise full electron pitch angle and energy distributions, instead of using the derived moments, we are successfully able to distinguish between multiple populations within what has historically been considered as one region, due to the lack of variation in the plasma moments (see table \ref{tab:parameters}) as well as the similarity in spatial location. Using 2D pitch angle and energy distributions also improves the time resolution of the plasma region classifications, due to a higher cadence in the spacecraft flux and counts data (e.g. 4 s resolution for the PEACE instrument) in comparison to the moments data (e.g. 8 s resolution for CIS moments and 16 s resolution for PEACE moments).

\section{Conclusion}

We present a novel machine learning method that characterises full particle distributions in order to classify different space plasma regimes. Our method uses autoencoders and subsequently principal component analysis to reduce the dimensionality of the 2D particle distributions to three dimensions. We then apply the mean shift algorithm to discover the number of populations in the dataset, followed by the agglomerative clustering algorithm to assign each data-point to a population. 

To illustrate the effectiveness of our method, we apply it to magnetotail electron data and compare our results to previous classifications, i.e. the ECLAT database, that utilises moments. With our method, we find multiple distinct electron populations within the plasma sheet, which previous studies have identified as one region (table \ref{tab:contingency_table}). These findings show that key features in particle distributions are not fully characterised by the plasma moments (e.g. table \ref{tab:parameters}), resulting in important distinctions between populations being overlooked. For example, we find two separate cold dense anisotropic populations in the plasma sheet (clusters 2 and 6), which are less abundant than the hotter and more isotropic plasma sheet populations. By using our clustering method to specify an exact list of times when populations like these are observed, we create a more comprehensive picture of their spatial distribution. Inherent time-dependencies may also contribute to our finding of multiple plasma sheet populations. Even in this case, our method is effective in characterising the evolution of particle populations, made possible by the high time resolution of our region classifications. In a follow up study, we will use this information to link the occurrence of these populations to other high-resolution spacecraft measurements in different plasma regions, in order to understand the physical processes driving changes in the less abundant particle populations. As an example analysis, our high resolution classifications of the observed anisotropic plasma sheet populations could be combined with previous theories on the sources of these populations \citep[e.g.][]{walsh_2013,Artemyev_2014}, to understand the relative contributions of particle outflows from distinct magnetospheric regions, such as the magnetosheath or ionosphere.

Comparisons between this original method and the previous classifications from ECLAT also show specific periods of disagreement (e.g. we classify a small number of ECLAT periods of plasma sheet as the plasma sheet boundary layer). This discrepancy shows that using the full 2D pitch angle and energy distributions, without requiring prior assumptions about magnetospheric plasma regions, may redefine the classifications of electron populations, along with our understanding of their plasma properties. Our method, which uses open-source and easily accessible machine learning techniques, can be used to better characterise any space plasma regime with sufficient in-situ observations. By not being constrained to a small number of parameters, this method allows for a more complete understanding of the interactions between various thermal and non-thermal populations. With increasingly large datasets being collected by multi-spacecraft missions, such as Cluster \citep{escoubet_2001} ($>$10\textsuperscript{9} full distributions in 20 years) and MMS \citep[Magnetospheric Multiscale Mission,][]{MMS}, similar methods would provide a useful tool to reduce the dimensionality of distributions, thereby optimising data retrieval on Earth. Furthermore, combining this method with large-scale survey data, such as NASA/GSFC's OMNI database, would allow users to isolate a specific population or plasma region for analysis of its properties.

\section*{Conflict of Interest Statement}

The authors declare that the research was conducted in the absence of any commercial or financial relationships that could be construed as a potential conflict of interest.

\section*{Author Contributions}

MRB developed the method described in the manuscript, tested it on the magnetotail data and wrote the manuscript. IJR was the lead supervisor who guided the direction of the project and provided insight at every stage. APW provided expertise on the magnetotail and the various populations that we observed, aiding the evaluation of our method. DV was also important in classifying the plasma regimes and provided insights into the physical processes governing electrons in space plasmas. AWS was key to the development of the method due to his expertise in machine learning. All co-authors made important contributions to the manuscript. 

\section*{Funding}

MRB is supported by a UCL Impact Studentship, joint funded by the ESA NPI programme. IJR the STFC Consolidated Grant ST/S000240/1 and the NERC grants NE/P017150/1, NE/P017185/1, NE/V002554/1, and NE/V002724/1. DV is supported by the STFC Consolidated Grant ST/S000240/1 and the STFC Ernest Rutherford Fellowship ST/P003826/1. AWS is supported by the STFC Consolidated Grant ST/S000240/1 and by NERC grants NE/P017150/1 and NE/V002724/1.

\section*{Acknowledgments}

We thank the Cluster instrument teams (PEACE, FGM, CIS, EFW) for the data used in this study, in particular the PEACE operations team at the Mullard Space Science Laboratory. We also acknowledge the European Union Framework 7 Programme, the ECLAT Project FP7 Grant no. 263325, and the ESA Cluster Science Archive.

\section*{Data Availability Statement}

The datasets analysed in this study can be found in the Cluster Science Archive (https://csa.esac.esa.int/csa-web/).

\bibliographystyle{frontiersinSCNS_ENG_HUMS} 
\bibliography{Paper_bib}


\end{document}